\documentclass[11pt]{article}

\usepackage[margin=0.9in]{geometry}
\usepackage{amsmath,amssymb,bm}
\usepackage{graphicx}
\usepackage{booktabs,array,multirow,tabularx}
\usepackage{caption}
\usepackage{float}
\usepackage{setspace}
\usepackage[hidelinks]{hyperref}
\usepackage{url}

\newcommand{\dd}{\mathrm{d}}
\newcommand{\R}{\mathbb{R}}
\newcommand{\vect}[1]{\bm{#1}}
\newcommand{\mat}[1]{\bm{#1}}
\newcommand{\norm}[1]{\left\lVert #1\right\rVert}
\newcommand{\Qdot}{\dot{Q}}
\newcommand{\mdot}{\dot{m}}

\newcommand{\lsp}{l_{sp}}

\newcommand{\uvbar}{\overline{u'v'}}

\newcolumntype{P}[1]{>{\raggedright\arraybackslash}p{#1}}
\newcommand{\figslot}[2]{%
\fbox{\begin{minipage}[c][#1][c]{0.90\linewidth}
\centering\small #2
\end{minipage}}}
\newcommand{\maybegraphics}[3][]{%
\IfFileExists{#2}{\includegraphics[#1]{#2}}{\figslot{#3}{\texttt{\detokenize{#2}}}}}

\begin{document}

\begin{center}
{\LARGE \bfseries Neural-Network and Reduced-order Modeling Workflows for AI-Driven CFD}\\[0.35cm]
{\large \bfseries Fast Response Surfaces, Reduced Dynamics and Jet in Cross-flow Examples}\\[0.55cm]
{\bfseries Kaku E. Eduku, Pavel P. Popov and Gustaaf Jacobs $^{*}$}\\[0.20cm]
Department of Aerospace Engineering, San Diego State University, San Diego, CA, USA\\
$^{*}$Corresponding author: gjacobs@sdsu.edu
\end{center}

\vspace{0.35cm}
\noindent \textbf{Abstract.}
Highly resolved computational fluid dynamics (CFD) simulations are essential tools in design, but they are too expensive to use as dense design-space samplers. Neural networks and data-driven approaches provide complementary tools for converting simulation campaigns into fast, interpretable and reusable models. This chapter presents a representative AI-driven CFD workflow using jet in cross-flow configurations. First, a reacting hydrogen jet in cross-flow is used to demonstrate a multilayer perceptron (MLP) response that maps injector spacing to three scalar design quantities: unburnt hydrogen throughput, wall heat transfer and bulk temperature concentration. A shape-preserving interpolation baseline is used to distinguish what is gained by the learned response from what can already be inferred directly from the CFD samples. The CFD samples show a non-monotonic spacing response, with the injector spacing of eight jet orifice diameters giving the lowest observed wall heat transfer and unburnt hydrogen throughput and the spacing of ten diameters giving the lowest bulk temperature. The MLP response identifies an intermediate-to-wide favorable region, with predicted heat-transfer and unburnt-hydrogen minima near $\mu\approx8.1$ and $\mu\approx7.8$, respectively. Leave-one-sample-out validation shows that the usefulness of the MLP depends strongly on the predicted quantity: the bulk temperature concentration is robust, with all errors below $4\%$ and an aggregate error of $1.46\%$, while heat transfer and unburnt hydrogen throughput have aggregate errors of $27.2\%$ and $21.7\%$. Second, Sparse Identification of Nonlinear Dynamics (SINDy) is employed as a reduced-order modeling framework for field-derived Reynolds-stress statistics. The parametric SINDy model provides compact field-level predictions at simulated and out-of-sample spacings, but its aggregate spacing-mean Reynolds-stress error is 6.7\% higher than the POD-bases reconstruction due to degradation at $\mu=8$ and $x/D=0$. The broader conclusion is that AI-driven CFD is not a single-model prescription. MLPs are effective for fast scalar responses, while POD--SINDy is better suited when transient reduced dynamics and field-derived statistics are central to the question.

\vspace{0.25cm}

\vspace{0.45cm}
\begin{center}
\textbf{Nomenclature}
\end{center}
{\small\begin{tabular}{P{0.15\linewidth}P{0.41\linewidth}}
$D$ & jet diameter\\
$\lsp$ & injector spacing\\
$\mu$ & nondimensional spacing\\
$\Qdot$ & area-averaged wall heat transfer rate\\
$\mdot$ & mass flow rate\\
$\vect a(t)$ & reduced coordinate vector\\
$\mat\Theta$ & SINDy library matrix\\
$\mat\Xi$ & sparse coefficient matrix\\
$\uvbar$ & Reynolds stress
\end{tabular}}

\vspace{0.65cm}
\begin{center}
\textbf{Abbreviations}
\end{center}
{\small\begin{tabular}{P{0.15\linewidth}P{0.41\linewidth}}
AI & artificial intelligence\\
CFD & computational fluid dynamics\\
JICF & jet in cross-flow\\
LOSO & leave-one-sample-out\\
MLP & multilayer perceptron\\
POD & proper orthogonal decomposition\\
ROM & reduced-order model\\
SINDy & Sparse Identification of Nonlinear Dynamics\\
QI & quantity of interest
\end{tabular}}

\section{Background/Introduction/Objective}

Highly resolved CFD remains indispensable for turbulent and reacting-flow analysis because it resolves spatial organization, unsteady transport and thermochemical coupling that are difficult to represent with lower-order engineering correlations. Its limitation is computational cost. A single well-resolved turbulent reacting-flow case requires sufficient mesh resolution, small time steps and detailed chemical mechanisms, making dense parametric sweeps expensive or impractical. This is particularly important for micromix combustors, where mixing and combustion occur at smaller scales \cite{Singh2025micromix}. The goal of AI in CFD studies is therefore to build computationally efficient models from highly resolved simulation data so that design questions can be asked more rapidly. 

Machine-learning methods enter CFD workflows in several distinct roles. They may accelerate expensive constitutive or chemistry evaluations, approximate closures, construct parameter-to-response surrogates, reconstruct fields from compressed or sparse representations, or advance reduced dynamical states. Reviews of machine learning in fluid mechanics and CFD emphasize that these roles have different data requirements and different relationships to the governing physics \cite{Brunton2020MLFluid,Vinuesa2022CFDML}. In combustion modeling, for example, neural networks have been used to approximate flamelet tables and chemistry-related source terms so that detailed thermochemical information can be retrieved with lower memory requirements or runtime cost \cite{Emami2012,Owoyele2019,Readshaw2021}. In aerodynamic design, neural-network surrogates and POD-neural ROMs have been used to predict aerodynamic or aerostructural responses over design spaces \cite{Park2013,Du2021}. Physics-informed neural networks have been proposed for differential-equation constrained regression and inverse problems \cite{Raissi2019}, while autoencoder and SINDy-based approaches have been used to represent high-dimensional flow evolution in lower-dimensional coordinates \cite{Hinton2006,Brunton2016,Fukami2021}. These examples illustrate a central point: AI-assisted CFD is not a single technique, but a family of tools whose usefulness depends on the data representation and the physical quantity of interest.

For an expensive CFD campaign, the most useful distinction is therefore not simply between ``classical'' and ``AI'' models, or between shallow and deep networks. It is between the mathematical objects that must be predicted. A map from a few design variables to integrated quantities of interest is a different problem from reconstruction of a spatial field, and both differ from propagating that field through time. Many-query aerodynamic studies make the same point from a surrogate-model perspective: model choice, sampling, validation and the intended use of the prediction must be considered together rather than independently \cite{Queipo2005Surrogate,Yondo2018Surrogate}. This chapter adopts that task-oriented view and illustrates how the representation required by a CFD question determines the model structure that is useful.

The highlighted example is a jet-in-cross-flow (JICF). For a reacting hydrogen JICF spacing study, the first objective is to predict scalar design quantities from a small set of highly resolved simulations. For that task, a multilayer perceptron (MLP) response is robust because the input and output are both low-dimensional. For the same reacting hydrogen JICF spacing dataset, the second objective is to obtain a Reynolds-stress statistic from reconstructed, time-dependent fields. For that task, a reduced-order dynamical model (SINDy) is employed because the output is not only a scalar function of a parameter, but also depends on the temporal evolution of the underlying flow field.

The objective is therefore to relate the CFD task to an appropriate data-driven representation. The chapter introduces neural-network architectures at the level needed for CFD model selection, presents an MLP surrogate workflow for scalar responses and employs parametric SINDy for time-dependent, field-level statistics.

\section{Approaches/Methods}

\subsection{Neural-network architectures in CFD workflows}

Neural networks are nonlinear function approximators whose architectures should be selected according to the structure of the CFD data. The most direct architecture is the fully connected feed-forward network, often referred to as a multilayer perceptron. It maps a finite input vector to a finite output vector through hidden layers and nonlinear activations \cite{Goodfellow2016}. This makes it well suited to parameter-to-response tasks, closure correlations, reduced-coordinate maps and low-dimensional design studies \cite{Park2013,Du2021,Ling2016,Duraisamy2019}. If the input is a design parameter such as injector spacing, Reynolds or Mach number, and the output is a scalar or a small vector of quantities of interest, the MLP is often an appropriate starting point.

Convolutional neural networks (CNNs) are more appropriate when the input has image-like spatial structure. They exploit local receptive fields and repeated filters, which makes them useful for field snapshots, flow visualization data, super-resolution, sensor-to-field reconstruction and learned low-dimensionalization \cite{LeCun2015,Fukami2021}. Recurrent neural networks (RNNs) are designed for sequence data and are therefore natural candidates for propagating temporal statistics or reduced-coordinate time series \cite{Hochreiter1997,Hasegawa2020CNNLSTM}.

Selection of the appropriate architecture is therefore fundamentally a question of representation. The relevant issue is what information must remain available in the model output. An integrated engineering quantity can be represented directly as a function of a small set of design or operating parameters, whereas prediction of a spatial field requires preservation of spatial structure. If the quantity of interest also depends on the evolution of that field, the model must additionally retain or represent temporal dynamics. Model complexity is consequently useful only insofar as it preserves information required by the CFD question.

Reduced-order dynamical models provide one route for retaining both spatial and temporal information without propagating the full CFD state. The field is first represented in a lower-dimensional coordinate system, after which the temporal evolution of those coordinates is modeled. Sparse Identification of Nonlinear Dynamics (SINDy), in particular, learns explicit evolution equations from reduced-coordinate data and therefore provides an interpretable dynamical representation.

The following sections use this distinction to motivate a parameter-to-response workflow first and a reduced-order framework workflow second. The MLP example asks how much design information can be extracted from a set of scalar CFD responses. The POD--SINDy example asks a harder question: whether enough spatial and temporal structure can be retained to reconstruct a field-derived statistic at both sampled and unsampled parameter values.

\subsection{CFD-derived quantities of interest}

The reacting hydrogen JICF configuration consists of a hydrogen jet issuing into a hot air cross-flow. The configuration uses a jet diameter of $D=1.27\,\mathrm{mm}$, a jet temperature of $\mathrm{T}_{jet}=300\,\mathrm{K}$, a cross-flow temperature of $\mathrm{T}_{\infty}=700\,\mathrm{K}$, an isothermal wall at $\mathrm{T}_{wall}=400\,\mathrm{K}$ and 1 atm pressure \cite{Eduku2026}. Periodic spanwise boundary conditions are used so that the spanwise width of the computational unit cell ($l_{sp}$) represents the spacing between adjacent injectors in an idealized row. The nondimensional design parameter is
\begin{equation}
    \mu=\frac{\lsp}{D}.
\end{equation}
The companion highly resolved CFD study \cite{Eduku2026} was conducted for cases of $\mu\in\{2,4,6,8,10\}$.

Three scalar quantities of interest are extracted from the simulated flow. The first is normalized unburnt hydrogen throughput,
\begin{equation}
    \frac{\mdot_{H_2,out}}{\mdot_{H_2,in}}
    =
    \frac{\int_{S_e}\rho Y_{H_2}\vect u\cdot\vect n\,\dd S}
         {\int_{S_i}\rho Y_{H_2}\vect u\cdot\vect n\,\dd S},
\end{equation}
which measures the fraction of injected hydrogen that leaves the analysis region without being consumed. The second is area-averaged wall heat transfer,
\begin{equation}
    \Qdot=\frac{1}{S_w}\int_{S_w}\vect q\cdot\vect n\,\dd S,
\end{equation}
which serves as a proxy for near-wall thermal loading. The third is a bulk temperature concentration metric,
\begin{equation}
    \frac{\norm{\mathrm{T}}_2}{\norm{\mathrm{T}}_1}
    =
    \frac{\left(V\int_{CV}\mathrm{T}^2\,\dd V\right)^{1/2}}
         {\int_{CV}\mathrm{T}\,\dd V},
\end{equation}
which characterizes how condensed or diffused the temperature field is. These quantities are deliberately scalar and design oriented. They are therefore suitable for a parameter-to-response surrogate approach to AI-driven CFD.

\subsection{Simple interpolation baseline for scalar data}

Before introducing a learned scalar surrogate, it is useful to establish what can be obtained from the CFD samples with a low-complexity interpolant. This provides a reference for deciding whether the flexibility of a neural model is actually useful for the quantity being predicted. Here, a separate piecewise cubic Hermite interpolating polynomial (PCHIP) is constructed for each time-averaged quantity of interest,
\begin{equation}
    \widehat{y}^{\mathrm{PCHIP}}_j(\mu)
    =
    \mathrm{PCHIP}\left\{(\mu_i,y_{j,i})\right\}_{i=1}^{n_s}(\mu),
\end{equation}
where $n_s=5$ is the number of simulated spacings and $y_{j,i}$ is the $j$-th scalar response at $\mu_i$. Shape-preserving piecewise cubic interpolation is attractive for a one-dimensional design variable because it follows local trends without imposing a single global polynomial, thereby reducing the risk of oscillations between samples \cite{FritschCarlson1980}. It is deterministic and requires no training procedure.

\subsection{MLP formulation and training}

The learning problem is a supervised regression task with injector spacing as the input and the QIs as outputs. The objective is therefore to approximate a low-dimensional nonlinear mapping from a single geometric control parameter to a small vector of integrated flow responses:
\begin{equation}
    \mu \longmapsto
    \left(\left\langle\frac{\mdot_{H_2,out}}{\mdot_{H_2,in}}\right\rangle,
    \langle\Qdot\rangle,
    \left\langle\frac{\norm{\mathrm{T}}_2}{\norm{\mathrm{T}}_1}\right\rangle\right).
\end{equation}
For $N_l$ hidden layers with widths $Q_1,\dots,Q_{N_{l}}$, the MLP hidden states are
\begin{equation}
    \vect q_{\ell}=f_{a,\ell}\left(W_{\ell}\vect q_{\ell-1}+\vect b_{\ell}\right),
    \qquad \ell=1,\ldots,N_l,
    \label{eqn:eqn34}
\end{equation}
where $\vect q_0$ is the spacing input, $W_{\ell}$ and $\vect b_{\ell}$ are trainable weights and biases, and $f_{a,\ell}$ is the layer activation. 
Equivalently, the $j$-th neuron in layer $\ell$ is
\begin{equation}
	q_{\ell,j}
	=
	f_{a,\ell}\!\left(
	\sum_{i=1}^{Q_{\ell-1}}
	W_{\ell,ji}q_{\ell-1,i}
	+
	b_{\ell,j}
	\right),
	\qquad
	j=1,\dots,Q_{\ell}.
	\label{eqn:eqn35}
\end{equation}
The output layer maps the final hidden representation to the predicted QIs
\begin{equation}
    \widehat{\vect y}=W_{out}\vect q_{N_l}+\vect b_{out}.
    \label{eqn:eqn36}
\end{equation}
Together, \eqref{eqn:eqn34}--\eqref{eqn:eqn36} define the functional relation of the parameter to the desired output. Pre-processing is the first step in the MLP training pipeline, here, the outputs are rescaled to [0,1] using the minima and maxima of the training dataset. Training is posed as minimization of the discrepancy between the MLP prediction and the simulated, true QIs, as such, optimization is driven by the mean-squared error
\begin{equation}
    \mathcal{L}^{MSE}=\frac{1}{n_s}\sum_{k=1}^{n_s}\left(\widehat{\vect y}_{(k)}-\vect y_{(k)}\right)^2.
\end{equation}
GELU activations are used because they provide smooth nonlinear mappings without prescribing the response shape \cite{Hendrycks2016,Hornik1989}. The model is trained with the Adam optimizer \cite{adam_opt} using a learning rate of $10^{-2}$ for 200 epochs; also, samples are shuffled each epoch. After training, the model is reverted to the best checkpoint (weights from the highest recorded accuracy) for prediction.

The small number of available simulations makes validation and model selection central. Multiple networks are trained from independent initializations, and a model is accepted only when the range-normalized root-mean-squared and pointwise errors satisfy
\begin{equation}
    \varepsilon_{\mu_{all}}
    =
    \frac{
    \sqrt{\frac{1}{n_s}\sum_{k=1}^{n_s}(\widehat y_k-y_k)^2}}
    {\max(y)-\min(y)} <0.05, \qquad
    \varepsilon_{\mu_i}
    =
    \frac{1}{3}
    \sum_{j=1}^{3}
    \frac{
    \left|
    \widehat{y}_{j}(\mu_i)-y_{j}(\mu_i)
    \right|}
    {y^{\max}_{j}-y^{\min}_{j}}<0.05 .
\end{equation}
The final reported response is the ensemble mean of twenty accepted models. This accepted-model ensemble reduces the dependence of the response curve on a single random initialization. 

\begin{table}[H]
\centering
\caption{MLP workflow used for the reacting hydrogen JICF spacing response.}
\label{tab:mlp_settings_short}
\begin{tabular}{@{}p{0.31\linewidth}p{0.64\linewidth}@{}}
\toprule
Item & Setting \\
\midrule
Input & Injector spacing, $\mu=\lsp/D$ \\
Outputs & Mean unburnt hydrogen throughput, mean area-averaged wall heat transfer and mean bulk temperature concentration \\
Architecture & MLP, 5 hidden layers with 16--32--32--24--12 neurons respectively \\
Training objective and optimizer & MSE on normalized outputs, Adam \cite{adam_opt}, learning rate of $10^{-2}$ \\
Training length & 200 epochs, five steps per epoch, randomly shuffled samples \\
Acceptance gate & $\varepsilon_{\mu_{\mathrm{all}}}<0.05$ and $\varepsilon_{\mu_i}<0.05$ for $\mu_i\in\{2,4,6,8,10\}$ \\
Ensemble size and band & 20 accepted models, with $\pm\,2$ ensemble standard deviations \\
Validation & Leave-one-sample-out tests \\
\bottomrule
\end{tabular}
\end{table}

\subsection{Reduced-coordinate dynamics from CFD fields}

SINDy learns explicit ordinary differential equations (ODEs) from time-resolved data \cite{Brunton2016}. For a state vector $\vect x(t)$,
\begin{equation}
    \frac{\dd}{\dd t}\vect x(t)=\vect f(\vect x(t)),
\end{equation}
SINDy approximates the unknown vector field by a sparse combination of candidate functions. With snapshot matrix $\mat X$, derivative matrix $\dot{\mat X}$ and library $\mat\Theta(\mat X)$, the sparse regression problem can be written as
\begin{equation}
    \min_{\mat\Xi}\;\norm{\dot{\mat X}-\mat\Theta(\mat X)\mat\Xi}_2^2+\lambda R(\mat\Xi),
\end{equation}
where $R(\mat\Xi)$ denotes a sparsity-promoting regularization term and $\mat\Xi$ contains the identified coefficients. The resulting model can be integrated forward in time and inspected term by term, making it attractive when interpretability and temporal rollout are important.

For high-dimensional CFD snapshots, SINDy is commonly applied after dimensionality reduction. If $\mat S(t)$ denotes a field snapshot such as velocity, vorticity or temperature, a POD approximation gives
\begin{equation}
    \mat S(t)\approx \sum_{i=1}^{r}a_i(t)\vect\phi_i,
\end{equation}
where $\vect\phi_i$ are spatial modes and $a_i(t)$ are temporal coefficients \cite{Chen2013}. SINDy is then trained on the reduced coordinate vector $\vect a(t)$. The learned ODEs can be integrated forward in time, after which the predicted coefficients are mapped through the retained POD bases to reconstruct the physical field. This workflow is more involved than an MLP response, but it provides a direct route from reduced dynamics to field-derived quantities. 

Projection-based model reduction is especially useful in many-query settings because the expensive field is represented by a small number of coordinates that can be evaluated repeatedly across parameters or time \cite{Benner2015ParametricMOR}. The compression, however, introduces its own approximation before any dynamical model is learned. POD is optimal for representing the supplied snapshots in a least-squares or energy sense, but a retained energy fraction is not itself a guarantee that every local feature or derived engineering statistic is equally accurate. A quantity such as Reynolds stress may depend on structures that are weak in the global energy ranking but important at a particular location.

This creates a useful error hierarchy for data-driven ROMs. The first question is whether the retained basis can reconstruct the field of interest when the true modal coefficients are supplied. Only after that test is the second question meaningful: how much additional error is introduced by predicting or propagating those coefficients? Separating these two levels prevents errors caused by an insufficient reduced basis from being attributed entirely to the temporal model. It also provides a clearer basis for deciding whether improvement should come from retaining more modes, changing the representation, or changing the dynamical identification procedure.

A POD--SINDy workflow therefore separates the problem into a spatial representation and a reduced temporal model: POD represents the field, SINDy evolves the reduced coordinates and the reconstructed field is post-processed into the desired statistic \cite{Fukami2021}.

\subsection{Parametric SINDy formulation}

Many CFD design studies depend on a parameter, such as injector spacing, Reynolds number or equivalence ratio. A non-parametric SINDy model trained at a single condition may identify a useful local vector field, but it does not directly describe how the reduced dynamics change across the design space. A parametric formulation introduces this dependence explicitly. 

Parametric sparse-identification methods treat the control parameter as part of the model-discovery problem rather than as a label for separate cases. Related formulations have been used to identify parameter-dependent PDEs with group sparsity, to construct reduced dynamics that can be queried at new parameter instances and to reduce the transient-data burden for parameterized model discovery \cite{Rudy2019ParametricPDE,Conti2023ParametricSINDy,Lemus2025ParameterizedSINDy}. For a reduced state $\vect a(t;\mu)\in\R^r$, the dynamics are written as
\begin{equation}
    \dot{\vect a}(t;\mu)=\vect F(\vect a(t;\mu),\mu).
\end{equation}
The implementation employed appends the design parameter to the library input,
\begin{equation}
    \widetilde{\vect a}(t;\mu)=
    \begin{bmatrix}
    \vect a(t;\mu) & \mu & \mu^2 & \cdots & \mu^p
    \end{bmatrix}^{T},
\end{equation}
so that
\begin{equation}
    \dot{\vect a}(t;\mu)\approx \mat\Theta\!\left(\widetilde{\vect a}(t;\mu)\right)\mat\Xi.
    \label{eq:param_sindy_appended}
\end{equation}
For a first-order polynomial parameter dependence, $p=1$, the appended-library treats $\mu$ as an additional library variable. If the state library is polynomial, this permits mixed state-parameter terms such as $\mu a_i$, $\mu a_i a_j$ or higher-order interactions depending on the selected polynomial order. The library may also be augmented with harmonic functions of the retained coordinates and appended parameter. In that case, the candidate library contains polynomial terms together with terms of the form
\begin{equation}
    \sin(k\widetilde a_j), \qquad \cos(k\widetilde a_j),
    \qquad k=1,\ldots,K,
\end{equation}
where $\widetilde a_j$ denotes an entry of the augmented vector $\widetilde{\vect a}$ and $K$ is the selected harmonic dimension. These harmonic terms provide additional candidates for representing oscillatory or phase-like reduced-coordinate behavior without changing the sparse-regression structure.

Parametric SINDy is attractive for CFD because it sits between a purely scalar surrogate and a full field solver, while retaining an explicit reduced dynamical model. Its performance remains sensitive to derivative estimates, library choice and rollout stability. Numerical differentiation converts noise or unresolved high-frequency content into the regression target; the candidate library determines which nonlinear interactions are available to be identified; and a sparse model that matches instantaneous derivatives can still accumulate substantial error when integrated over a long horizon. Modern SINDy software and methodological developments therefore emphasize differentiation choices, regularized sparse regression and validation of identified models beyond the regression residual alone \cite{Kaptanoglu2022PySINDy}.

Parametric SINDy consequently provides a compact way to describe how reduced dynamics vary across operating conditions and is a natural extension when the response of interest is tied to time evolution rather than direct parameter-to-scalar regression.

\subsection{Field-reconstructed Reynolds stress}
The representative SINDy application is drawn from the same reacting hydrogen jet in cross-flow spacing dataset used for the MLP response. In this case, the objective is to compute Reynolds stress; as such, the velocity field is reduced using POD, the retained coefficients are evolved using the appended-variable parametric SINDy model and the metric is obtained from the field reconstruction. The Reynolds decomposition is $\vect u = \overline{\vect u}+\vect u'$, with the Reynolds stress given by $\uvbar$.

For this dataset, POD is applied to the velocity snapshots and $r=5$ velocity modes are retained. The appended-library formulation is employed with spacing mapped to the normalized parameter $p(\mu)$ so that the augmented state library used is
\begin{equation}
    \widetilde{\vect a}(t;\mu)
    =
    \begin{bmatrix}
    a_1(t;\mu) & a_2(t;\mu) & a_3(t;\mu) & a_4(t;\mu) & a_5(t;\mu) & p(\mu)
    \end{bmatrix}^{T}.
\end{equation}

The time step used for the derivative calculation and model rollout is $\Delta t_D =
6\times 10^{-6} \mathrm{s}/D \approx 0.0048\ \mathrm{s/m}.$ Time derivatives of the modal coefficients are estimated with a third-order Savitzky--Golay \cite{SavitzkyGolay} differentiating filter using a 21-point window. The candidate library uses a cubic polynomial basis in the augmented variables together with harmonic terms up to $K=8$. Before sparse regression, the library columns are normalized according to
\begin{equation}
    \mat \Theta
    =
    \mat \Theta_0 \mat C^{-1},
    \qquad
    C_{\ell\ell}
    =
    \left\|
    \mat \Theta_{0,:,\,\ell}
    \right\|_2
    +
    10^{-12}.
\end{equation}
The sparse coefficient matrix represented in this chapter is obtained using sequential thresholded ridge regression with ridge parameter $\rho=1.3\times10^{-1}$ and sparsity threshold $\lambda=2.23\times10^{-1}$.

The sparsity threshold $\lambda$ and ridge parameter $\rho$ are selected through a grid search over candidate $(\lambda,\rho)$ pairs. For each pair, a sequential thresholded ridge-regression model is first fit in the column-normalized library. The ridge parameter then regularizes active least-squares refits, while the sparsity threshold removes terms according to their root-mean-square contribution to each reduced equation,
\begin{equation}
    c_{\ell j}
    =
    \left[
    \frac{1}{N}
    \sum_{n=1}^{N}
    \left(
    \Theta_{N,n\ell}\Xi_{\ell j}
    \right)^2
    \right]^{1/2},
\end{equation}
where $\Theta_N$ is the normalized library, $\Xi_{\ell j}$ is the coefficient of the $\ell$-th candidate term in the $j$-th reduced equation and $N$ is the number of differentiated training samples. Candidate models are first screened by derivative-space accuracy, correlation, amplitude and linearized-stability diagnostics. A shortlisted subset is then integrated over the training spacings, and the final $(\lambda,\rho)$ pair is selected from the candidates that produce stable reduced-coordinate rollouts.

For rollout selection, each shortlisted candidate model is integrated using fixed-step fourth-order Runge--Kutta (RK4) integration and compared with the corresponding POD-coefficient trajectories. The fixed-step scheme provides a consistent and computationally efficient basis for comparing the large number of candidate models during the search. The rollout score is a weighted combination of the normalized mean-square trajectory error, a correlation loss, an amplitude loss and a linearized-stability loss,
\begin{equation}
    \mathcal{J}_{\mathrm{roll}}
    =
    w_{\mathrm{mse}}\mathcal{L}_{\mathrm{mse}}
    +
    w_{\mathrm{corr}}\mathcal{L}_{\mathrm{corr}}
    +
    w_{\mathrm{amp}}\mathcal{L}_{\mathrm{amp}}
    +
    w_{\mathrm{eig}}\mathcal{L}_{\mathrm{eig}} .
\end{equation}
The normalized trajectory loss is
\begin{equation}
    \mathcal{L}_{\mathrm{mse}}
    =
    \frac{
    \left\langle
    \left(
    \widehat{a}_{j}(t;\mu)-a_{j}(t;\mu)
    \right)^2
    \right\rangle_{\mu,t,j}}
    {
    \left\langle
    a_{j}(t;\mu)^2
    \right\rangle_{\mu,t,j}
    +\varepsilon
    },
\end{equation}
where $a_j(t;\mu)$ and $\widehat{a}_j(t;\mu)$ are the true and SINDy-predicted POD coefficients, respectively. The correlation loss penalizes phase and shape disagreement,
\begin{equation}
    \mathcal{L}_{\mathrm{corr}}
    =
    \frac{1}{r}
    \sum_{j=1}^{r}
    \left(
    1-\left|\rho_j\right|
    \right),
    \qquad
    \rho_j
    =
    \frac{
    \sum_{\mu,t}
    \left(
    \widehat{a}_{j}-\overline{\widehat{a}}_{j}
    \right)
    \left(
    a_j-\overline{a}_j
    \right)}
    {
    \left[
    \sum_{\mu,t}
    \left(
    \widehat{a}_{j}-\overline{\widehat{a}}_{j}
    \right)^2
    \right]^{1/2}
    \left[
    \sum_{\mu,t}
    \left(
    a_j-\overline{a}_j
    \right)^2
    \right]^{1/2}
    +\varepsilon
    } .
\end{equation}
The amplitude loss penalizes trajectories with incorrect modal energy,
\begin{equation}
    \mathcal{L}_{\mathrm{amp}}
    =
    \frac{1}{r}
    \sum_{j=1}^{r}
    \left[
    \log
    \left(
    \frac{
    \mathrm{rms}\left(\widehat{a}_j\right)}
    {
    \mathrm{rms}\left(a_j\right)+\varepsilon}
    \right)
    \right]^2 ,
\end{equation}
and the stability loss penalizes unstable linearized dynamics,
\begin{equation}
    \mathcal{L}_{\mathrm{eig}}
    =
    \left[
    \max
    \left(
    0,
    \sigma_{\max}-\epsilon_{\mathrm{stable}}
    \right)
    \right]^2
    +
    \eta_{\mathrm{eig}}\sigma_{\max}^{2},
    \qquad
    \sigma_{\max}
    =
    \max_{\mu}
    \max
    \operatorname{Re}
    \left[
    \lambda_i\left(A(\mu)\right)
    \right],
\end{equation}
where $A(\mu)$ is the affine linear block extracted from the sparse reduced-order model at spacing $\mu$. Rollouts that become non-finite, exceed prescribed absolute or relative amplitude limits, or exhibit step-to-step norm blow-up are rejected. Among the remaining candidates, the selected SINDy model is the one with the smallest $\mathcal{J}_{\mathrm{roll}}$.

\begin{figure}[H]
\centering
\maybegraphics[width=0.98\linewidth]{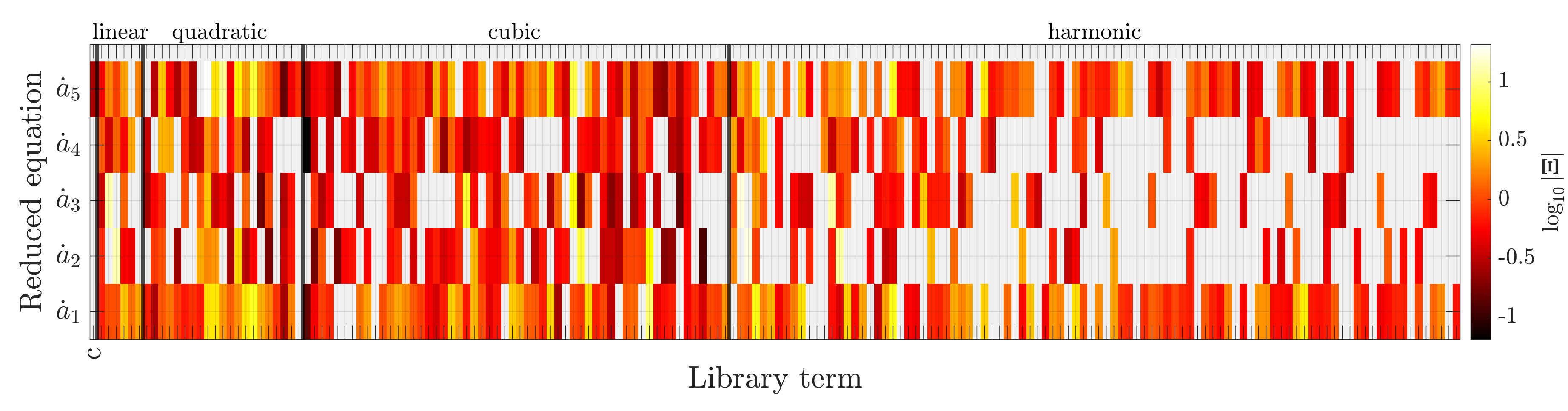}{2.25in}
\caption{Representative sparse coefficient matrix for the appended-variable parametric SINDy model. The library employs five retained velocity coefficients, a cubic polynomial state library with a $K=8$ harmonic dimension and a first-order appended spacing parameter $\mu$. The heat map identifies the active library terms and the relative coefficient magnitude in each of the five reduced equations.}
\label{fig:parametric_library}
\end{figure}

After model selection, the identified ODEs used for the reported field reconstructions are re-integrated using the MATLAB \texttt{ode45} solver with a tolerance of $10^{-8}$. A solution is retained only if it supplies at least one steady-window length (matching the training data length). The resulting sparse coefficient matrix contains five reduced equations, one for each retained velocity coefficient. A heat-map representation of $\mat\Xi$ is included in figure \ref{fig:parametric_library} to show which candidate terms remain active after sparse regression and which equations contain the strongest identified couplings.

The identified coefficient matrix therefore defines the reduced dynamical system. The parametric model can also be evaluated at out-of-sample spacings. For this use case, the initial reduced coefficients at $\mu^*$ are obtained by PCHIP interpolation across the simulated spacing cases, after which the appended-variable SINDy model is integrated at the target parameter value. These predictions provide field-level trends at design values that were not simulated directly. A companion Python/JAX implementation of the parametric SINDy grid-search and rollout workflow, demonstrated using a synthetic five-state benchmark, is publicly available \cite{parametric-sindy-jax}.

\begin{table}[H]
\centering
\caption{Parametric SINDy workflow used for the reconstructed Reynolds stress.}
\label{tab:sindy_settings}
\begin{tabular}{@{}p{0.28\linewidth}p{0.68\linewidth}@{}}
\toprule
Item & Setting \\
\midrule
Training spacings & $\mu=2,4,6,8,10$ \\
Out-of-sample spacings & $\mu^{*}=3,5,7,9$ \\
Parameter scaling & $p(\mu)=2(\mu-\mu_{\min})/(\mu_{\max}-\mu_{\min})-1$ \\
Retained POD modes & Five modes, capturing $92.42\%$ of POD energy \\
Augmented variables & Five modal coefficients plus normalized spacing parameter $p(\mu)$ \\
Derivative estimate & Third-order Savitzky--Golay \cite{SavitzkyGolay}, 21-point window \\
Library & Cubic polynomial terms and harmonic terms through $K=8$ \\
Sparse regression & Sequential thresholded ridge regression, $\lambda=2.23\times10^{-1}$, $\rho=1.3\times10^{-1}$ \\
Grid-search parameters & $w_{\rm mse}=0.45$, $w_{\rm corr}=0.75$, $w_{\rm amp}=0.1$, $w_{\rm eig}=10^{-3}$, $\epsilon_{\mathrm{stable}}=1.0$ \\
Out-of-sample initialization & PCHIP interpolation of initial POD coefficients \\
\bottomrule
\end{tabular}
\end{table}

\section{Results and Discussion}
\label{sec:results_discussion}

\subsection{Representative MLP response and validation}

\begin{table}[H]
\centering
\caption{Time-averaged design quantities for the simulated spacing cases, from the highly resolved CFD study of Eduku et al. \cite{Eduku2026}.}
\label{tab:qoi_compact}
\small
\begin{tabular}{@{}ccccc@{}}
\toprule
Case & $\mu$ & $\left\langle\mdot_{H_2,out}/\mdot_{H_2,in}\right\rangle$ & $\langle\Qdot\rangle$ (MW/m$^2$) & $\left\langle\norm{\mathrm{T}}_2/\norm{\mathrm{T}}_1\right\rangle$ \\
\midrule
S2  & 2  & 0.789 & 0.575 & 1.147 \\
S4  & 4  & 0.657 & 0.468 & 1.121 \\
S6  & 6  & 0.668 & 0.594 & 1.084 \\
S8  & 8  & 0.542 & 0.392 & 1.087 \\
S10 & 10 & 0.679 & 0.439 & 1.077 \\
\bottomrule
\end{tabular}
\end{table}

Table \ref{tab:qoi_compact} summarizes the CFD-derived quantities used here to construct the scalar response models. The response is non-monotonic in spacing. The narrowest case has the largest unburnt hydrogen throughput and the strongest bulk temperature concentration. The S8 case gives the lowest observed wall heat transfer and hydrogen throughput, while the S10 case lowers the temperature concentration metric further but increases the other two quantities relative to S8. This already indicates that spacing selection is a trade-off rather than a monotonic optimization problem.

\begin{figure}[t]
\centering
\maybegraphics[width=0.9\linewidth]{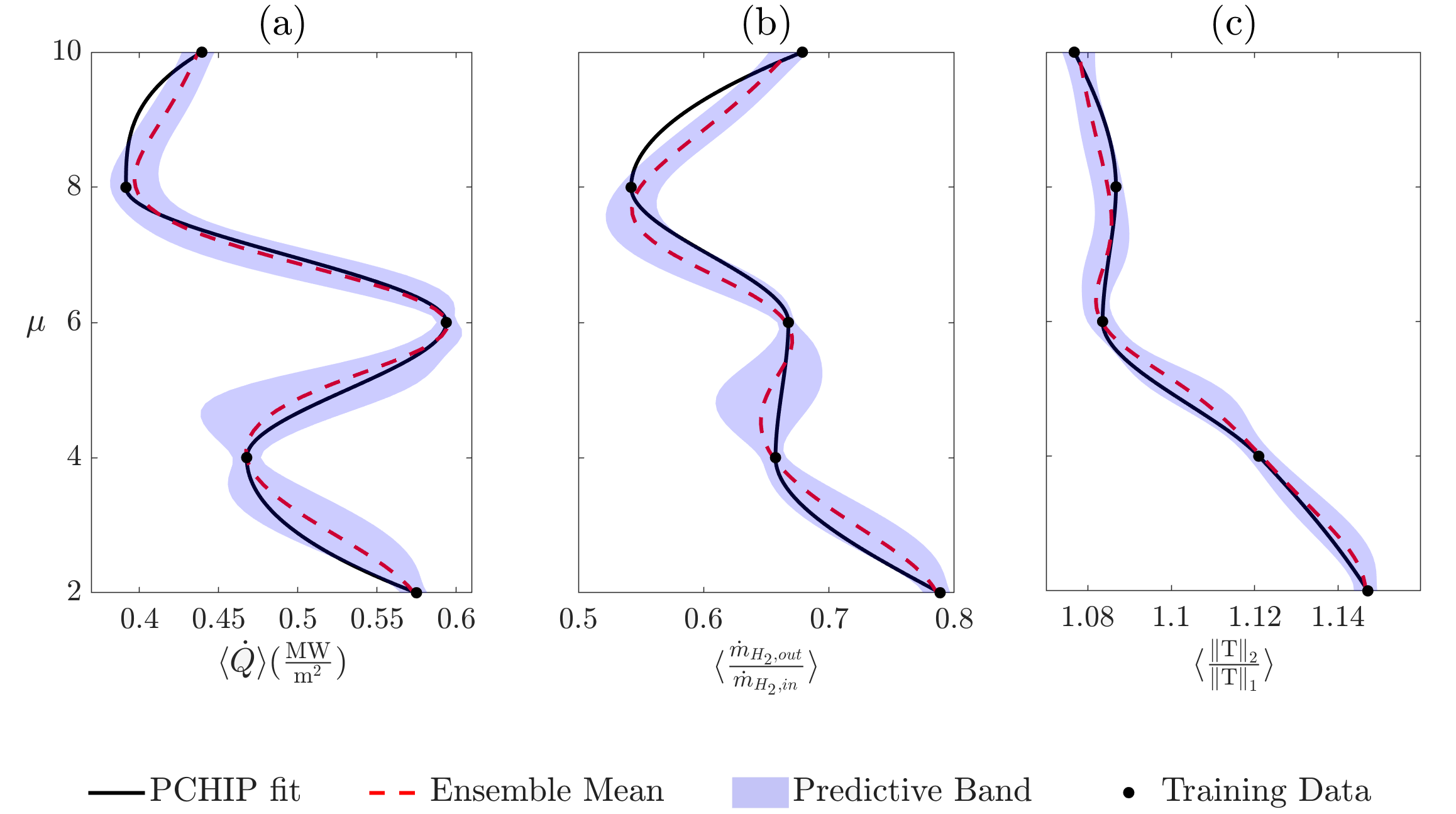}{1.85in}
\caption{PCHIP and MLP responses for the reacting hydrogen JICF spacing study. The MLP ensemble mean and corresponding spread are shown together with the PCHIP representation for the three scalar quantities of interest.}
\label{fig:mlp_response_slot}
\end{figure}

Figure \ref{fig:mlp_response_slot} highlights the PCHIP representation and MLP response depicting the ensemble mean and corresponding predictive band. This band is the ensemble-spread band, $y^{\pm}_{j}(\mu) = \overline{y}_{j}(\mu) \pm 2s_j(\mu)$ where $\overline{y}_{j}$ is the mean and $2s_j$ is the standard deviation, and should therefore be interpreted as spread across accepted initializations.

The scalar responses identify the same practical design message as the CFD quantities. The favorable region lies toward the intermediate and wide end of the sampled spacing range. The predicted heat-transfer minimum occurs near $\mu\approx8.1$ (figure \ref{fig:mlp_response_slot}a) for the MLP while the PCHIP minimum is at $\mu=8$. The predicted unburnt-hydrogen minimum occurs near $\mu\approx7.8$ for the MLP and $\mu=8$ for the PCHIP fit in figure \ref{fig:mlp_response_slot}b. The temperature concentration metric decreases from the narrow end to the wide end of the sampled range. The responses generated by both surrogates are similar with the most obvious differences being heat transfer and unburnt hydrogen in the interval $8\leq \mu \leq 10$ where the PCHIP solution exceeds the bounds of the MLP predictive band.

However, the value of the MLP must be interpreted in conjunction with out-of-sample validation and with the simpler interpolation reference. Leave-one-sample-out (LOSO) validation is performed by omitting one simulated spacing from training and comparing the inverse-scaled prediction with the omitted CFD value. Table \ref{tab:mlp_loso_compact} reports LOSO relative errors for both surrogates. The interior cases $\mu=4,\ 6,\ 8$ therefore test interpolation, whereas omission of $\mu=2$ or 10 forces one-sided extrapolation beyond the retained sample range for the PCHIP surrogate. The distinction matters most for wall heat transfer: PCHIP is competitive with the MLP at some interior points but produces negative extrapolated heat-transfer values at both endpoint holdouts, leading to errors above $100\%$. Shape preservation between samples does not imply physically admissible extrapolation beyond them; consequently, the interpolation aggregate is $\sim34\%$ with the full aggregate error being 65.4\%.

\begin{table}[H]
\centering
\caption{LOSO relative error (\%) of the PCHIP interpolation baseline and MLP response surface using the scalar values in Table \ref{tab:qoi_compact}. Endpoint PCHIP cases are extrapolations.}
\label{tab:mlp_loso_compact}
\scriptsize
\begin{tabular}{@{}c cc cc cc@{}}
\toprule
& \multicolumn{2}{c}{$\langle\Qdot\rangle$}
& \multicolumn{2}{c}{$\left\langle\mdot_{H_2,out}/\mdot_{H_2,in}\right\rangle$}
& \multicolumn{2}{c}{$\left\langle\norm{\mathrm{T}}_2/\norm{\mathrm{T}}_1\right\rangle$} \\
\cmidrule(lr){2-3}\cmidrule(lr){4-5}\cmidrule(l){6-7}
Omitted case & PCHIP & MLP & PCHIP & MLP & PCHIP & MLP \\
\midrule
S2  & 110.8 & 23.0 & 26.5 & 21.7 & 1.48 & 3.60 \\
S4  & 26.4  & 27.6 & 13.5 & 17.2 & 1.71 & 0.50 \\
S6  & 30.1  & 32.3 & 13.4 & 18.6 & 1.64 & 1.50 \\
S8  & 45.4  & 27.3 & 24.5 & 24.9 & 0.76 & 0.70 \\
S10 & 114.1 & 25.7 & 42.0 & 26.1 & 2.88 & 1.00 \\
\midrule
Aggregate & 65.4 & 27.2 & 24.0 & 21.7 & 1.69 & 1.46 \\
\bottomrule
\end{tabular}
\end{table}

The comparison also shows that surrogate suitability is quantity dependent. The bulk temperature concentration is represented well by both approaches: all PCHIP errors remain below $3\%$ and all MLP errors remain below $4\%$. Unburnt hydrogen throughput is more difficult, and the two approaches have comparable aggregate performance, with the MLP providing a modest improvement. Wall heat transfer is the most demanding response; here the MLP avoids the nonphysical endpoint behavior of the interpolant but still has an aggregate error of $27.2\%$. The MLP is therefore useful as a screening tool; it converts expensive cases into a continuous response and suggests where additional simulations should be focused. However, the LOSO results show that its predictive accuracy depends on the quantity being modeled.

A direct response surface is appropriate when the objective is a fast map from a design parameter to a scalar quantity. When the desired quantity instead depends on dynamics that must also be represented, the model must retain the route by which that quantity is produced rather than only its final integrated value.

\subsection{Representative SINDy application: Reynolds stress}

\begin{figure}[H]
\centering
\maybegraphics[width=0.98\linewidth]{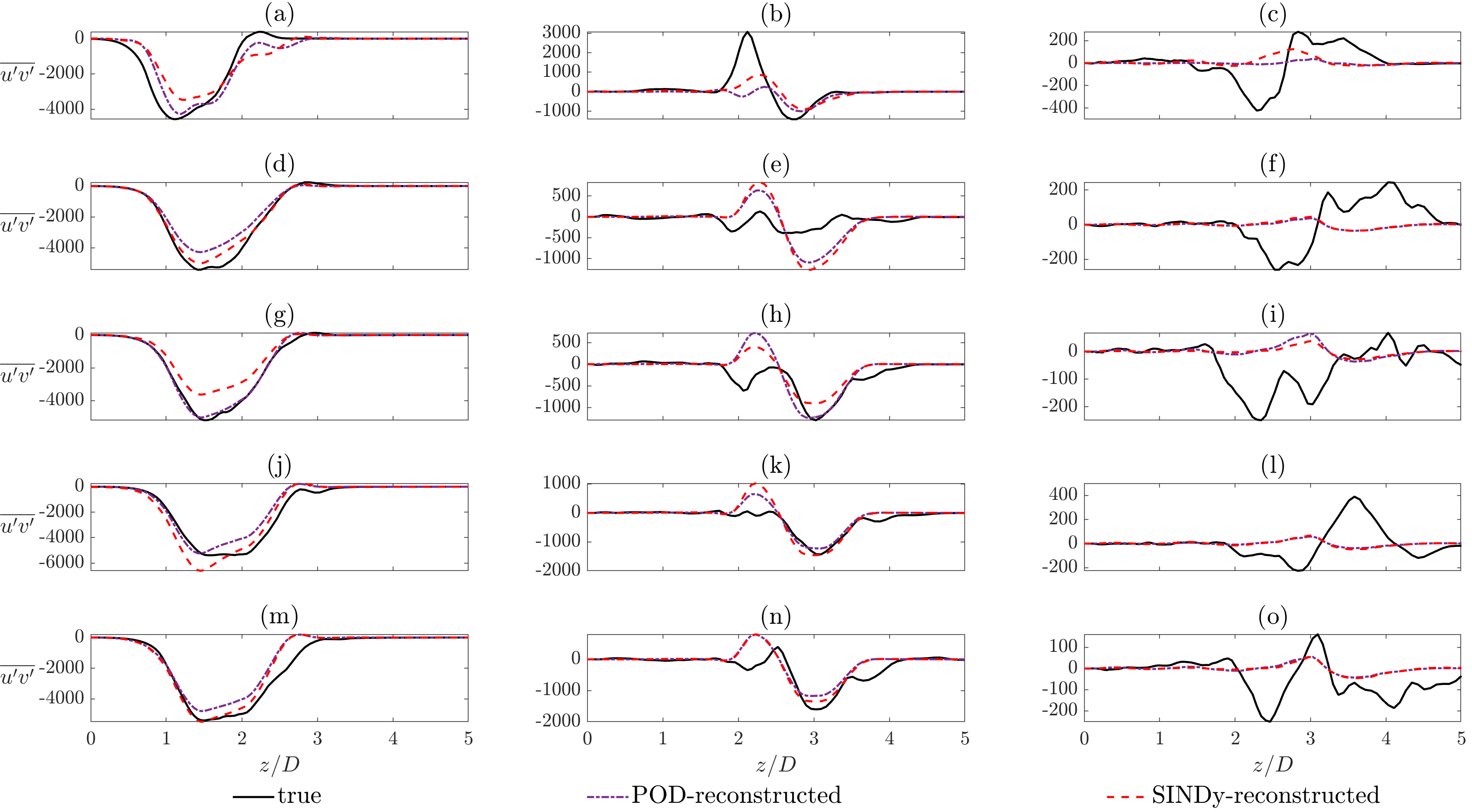}{3.00in}
\caption{Centerplane Reynolds-stress comparison for the reacting hydrogen jet in cross-flow cases. From left to right, columns correspond to $x/D=0$, $x/D=1$ and $x/D=2$, while rows correspond to $\mu=2,4,6,8$ and 10 from top to bottom. The profiles compare the true $\uvbar$ (black solid lines), the POD reconstruction using true retained coefficients (purple dash-dot lines) and the reconstruction obtained from parametric SINDy-propagated coefficients (red dashed lines).}
\label{fig:uvbar_comp}
\end{figure}

Figure \ref{fig:uvbar_comp} compares the true Reynolds-stress profiles, POD reconstruction using the true retained coefficients and SINDy reconstruction using the propagated coefficients. Table \ref{tab:parametric_reynolds_sindy} reports the range-normalized RMSE for both the POD reconstruction and the SINDy reconstruction.

The reduced representation itself introduces error especially for $x/D=2$ where neither the local minima nor maxima are captured for all $\mu$ cases. This means that five retained modes, although capturing $92.42\%$ of the POD energy, are insufficient to fully recover the downstream Reynolds stress profiles. This is an important ROM distinction: retained energy describes how well the snapshot ensemble is represented globally, whereas the engineering quantity may emphasize localized correlations or extrema that are not dominant in that energy measure. The POD-only column in Table \ref{tab:parametric_reynolds_sindy} therefore establishes the representation error that exists before the reduced coordinates are predicted dynamically.

For $x/D=0$, the reduced basis is most accurate with an aggregate nRMSE of 0.088. Here, the SINDy propagated solution experiences the greatest departure from the POD reconstruction with an increase of 19.3\% to the aggregate nRMSE, powered primarily by the overprediction of the maximum for $\mu=4$ and 8. Across all spacings and locations, the aggregate spacing-mean nRMSE for the SINDy solution is 6.7\% greater than the POD-bases reconstruction. Thus, the SINDy rollout adds temporal-modeling error to the retained POD bases in the aggregate, although localized improvements occur at selected spacings and planes. The only spacing-mean improvement occurs at $\mu=10$, where the SINDy error is 0.139, a 7.1\% decrease from the POD reconstruction. The largest SINDy degradation occurs at $\mu=8$, with a 16.7\% spacing-mean error increase.

\begin{figure}[t]
\centering
\maybegraphics[width=0.90\linewidth]{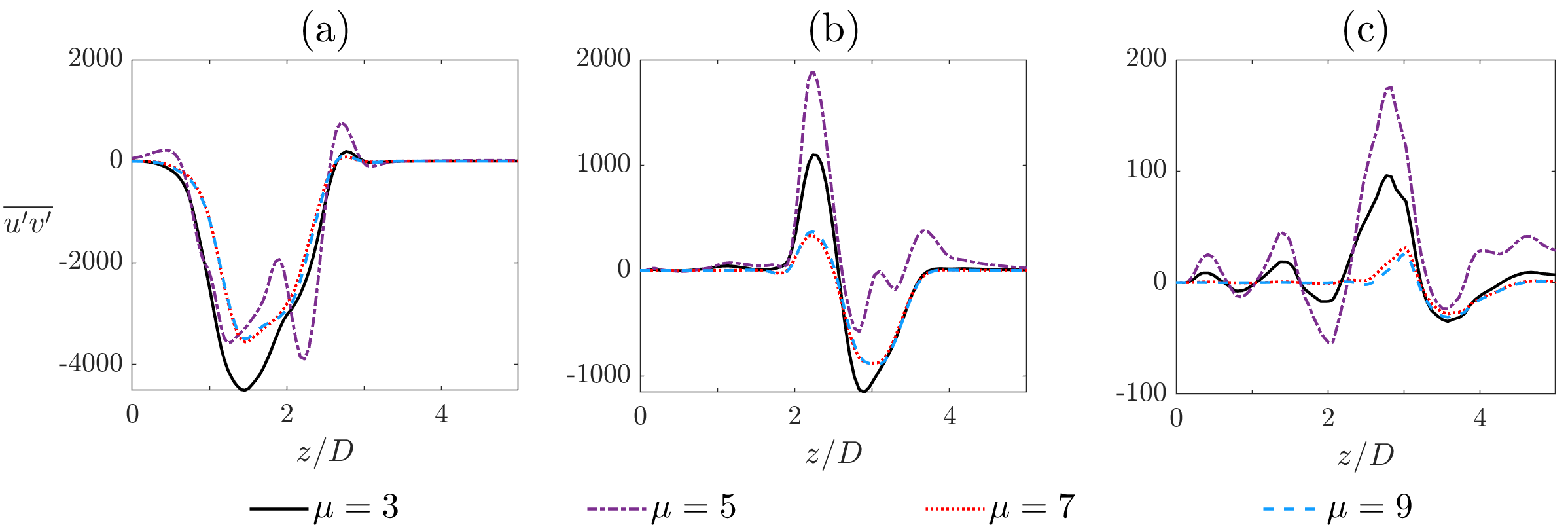}{2.20in}
\caption{Parametric SINDy Reynolds-stress predictions at out-of-sample spacing values. The profiles are evaluated at (a) $x/D=0$, (b) $x/D=1$ and (c) $x/D=2$. The solution is obtained from an appended-variable parametric library and PCHIP-interpolated initial reduced coefficients for $\mu=3,5,7$ and 9.}
\label{fig:uvbar_pred}
\end{figure}

Figure \ref{fig:uvbar_pred} demonstrates parametric evaluation at unsimulated spacing values of $\mu=3,5,7$ and 9. $\mu=3$ exhibits the local minima for $x/D=0$ and 1, however, at $x/D=2$, the spacing with the minimum is $\mu=5$. Also, $\mu=5$ consistently exhibits higher peaks at all $x/D$ locations. $\mu=7$ and 9 by contrast, are very similar in trend and magnitudes at all $x/D$ locations matching the trends exhibited by cases $\mu \geq 6$.

The Reynolds-stress example, therefore, demonstrates another use of machine learning. The objective is no longer a fast parameter-to-response map; it is now to obtain a field-derived statistic at simulated and out-of-sample design values. In this setting, SINDy provides a compact reduced-dynamics model that performs the spatio-temporal task through POD reduced coordinates, sparse temporal evolution and field reconstruction.

\begin{table}[H]
\centering
\caption{Range-normalized RMSE of centerplane Reynolds stress, $\uvbar$, from POD-only and parametric SINDy-reconstructed velocity fields.}
\label{tab:parametric_reynolds_sindy}
\scriptsize
\begin{tabular}{@{}c cc cc cc cc@{}}
\toprule
\multirow{2}{*}{$\mu$}
& \multicolumn{2}{c}{$x/D=0$}
& \multicolumn{2}{c}{$x/D=1$}
& \multicolumn{2}{c}{$x/D=2$}
& \multicolumn{2}{c}{Spacing mean} \\
\cmidrule(lr){2-3}
\cmidrule(lr){4-5}
\cmidrule(lr){6-7}
\cmidrule(l){8-9}
& POD & SINDy & POD & SINDy & POD & SINDy & POD & SINDy \\
\midrule
2  & 0.110 & 0.157 & 0.151 & 0.120 & 0.190 & 0.203 & 0.150 & 0.160 \\
4  & 0.099 & 0.046 & 0.568 & 0.696 & 0.222 & 0.225 & 0.296 & 0.322 \\
6  & 0.023 & 0.123 & 0.216 & 0.174 & 0.292 & 0.279 & 0.177 & 0.192 \\
8  & 0.100 & 0.112 & 0.118 & 0.172 & 0.202 & 0.206 & 0.140 & 0.163 \\
10 & 0.108 & 0.085 & 0.150 & 0.144 & 0.190 & 0.187 & 0.149 & 0.139 \\
\midrule
Aggregate & 0.088 & 0.105 & 0.241 & 0.261 & 0.219 & 0.220 & 0.183 & 0.195 \\
\bottomrule
\end{tabular}
\end{table}

\subsection{Choosing the right machine-learning tool}

The preceding examples demonstrate that the appropriate machine-learning tool should be selected from the CFD question rather than from model complexity alone. Table \ref{tab:tool_choice} summarizes the distinction.

\begin{table}[H]
\centering
\caption{Practical model choice for AI-driven CFD workflows.}
\label{tab:tool_choice}
\small
\begin{tabular}{@{}P{0.20\linewidth}P{0.25\linewidth}P{0.25\linewidth}P{0.20\linewidth}@{}}
\toprule
Tool & Best suited task & Main strength & Main weakness \\
\midrule
MLP & Fast scalar maps from design parameters to QIs & Simple and effective as a screening tool & Does not retain spatial or temporal field structure; accuracy is quantity dependent \\
POD--SINDy & Time evolution of reduced CFD fields and derived statistics & Retains dominant spatial structures with explicit reduced dynamics & Basis truncation and identified dynamics both contribute error \\
Parametric SINDy & Reduced dynamics across design parameters & Describes how reduced dynamics vary with operating conditions & Sensitive to derivative quality, library choice and rollout stability\\
\bottomrule
\end{tabular}
\end{table}

For the hydrogen spacing study, the MLP provides a direct parameter-to-response representation of the scalar design quantities. The PCHIP baseline gives a low-complexity reference for the same task, while the LOSO comparison shows that the usefulness of either scalar representation depends on the quantity being modeled. The MLP response is especially effective for the bulk temperature concentration metric and provides a useful screening representation for the more difficult heat-transfer and unburnt-hydrogen responses.

The Reynolds-stress problem requires a different representation because the desired statistic is obtained from an evolving velocity field. POD supplies the spatial reduction and SINDy supplies the reduced temporal dynamics. The comparison between POD-only and POD--SINDy reconstructions further separates error associated with the retained spatial basis from the additional error introduced by the identified dynamics. This distinction is useful whenever a reduced dynamical model is used to recover a field-derived quantity.

The appropriate model is therefore determined primarily by the mathematical object that must be predicted and the information that must be preserved. A direct surrogate is sufficient when the objective is a scalar parameter-to-response map; a reduced dynamical model is appropriate when temporal evolution and field reconstruction are central to the quantity of interest. The two workflows used here are complementary examples of that task-oriented model selection rather than steps in a hierarchy of increasing model complexity.

\section{Conclusion}

This chapter presented a representative AI-driven CFD workflow using neural-network response surfaces and reduced-order machine-learning models. The MLP example showed how a set of reacting hydrogen jet in cross-flow simulations can be converted into a fast spacing-response surrogate. The response identifies an intermediate-to-wide spacing region near $\mu\approx8$ as favorable among the simulated cases, but LOSO validation shows that heat transfer and unburnt hydrogen throughput are substantially harder to predict than the bulk temperature concentration metric.

SINDy was introduced as the more structured option when scalar regression is not sufficient. In the Reynolds-stress example, an appended-variable parametric SINDy model was used to evolve reduced velocity coefficients and reconstruct centerplane Reynolds-stress profiles. The best SINDy case occurs at $\mu=10$, while the strongest degradation appears at $\mu=8$ and in the $x/D=0$ profiles. The trained model can also be evaluated at out-of-sample spacing values using interpolated initial reduced coordinates, producing field-level predictions at design points that were not simulated directly.

The broader conclusion is that effective AI-driven CFD depends on matching the model representation to the use case. PCHIP provides a low-complexity reference for the scalar response, while the MLP provides a learned parameter-to-response map whose accuracy varies with the quantity of interest. POD--SINDy provides a different capability by retaining reduced temporal dynamics from which field-derived statistics can be reconstructed. Its performance reflects both the accuracy of the retained spatial basis and the additional error introduced by the identified dynamics.

Used together, these tools provide complementary insight into CFD-driven design: one summarizes the scalar design response, while the other connects unsimulated design values to reduced temporal dynamics and time-dependent flow statistics. More generally, the appropriate data-driven model is determined by what information the CFD question requires the surrogate to preserve, particularly whether the desired prediction is a scalar response or depends on spatial structure and temporal evolution.

\section{Acknowledgements}
The authors gratefully acknowledge support from the National Energy Technology Laboratory’s University Turbine System Research Program under grant number  DE-FE0032080. The technical input and feedback of Solar Turbines Incorporated is also very much appreciated.

\end{document}